\documentclass[]{article}
\usepackage[final]{nips_2018}
\usepackage[utf8]{inputenc} % allow utf-8 input
\usepackage[T1]{fontenc}    % use 8-bit T1 fonts
\usepackage{hyperref}       % hyperlinks
\usepackage{url}            % simple URL typesetting
\usepackage{booktabs}       % professional-quality tables
\usepackage{amsfonts}       % blackboard math symbols
\usepackage{nicefrac}       % compact symbols for 1/2, etc.
\usepackage{microtype}      % microtypography
\usepackage{amsmath}
\usepackage[pdftex]{graphicx}
\usepackage{subcaption}
\usepackage[ruled, vlined]{algorithm2e}

\title{Incorporating Biological Knowledge with Factor Graph Neural Network for Interpretable Deep Learning}

\author{
	Tianle Ma \\
	Department of Computer Science\\
	University at Buffalo\\
	Buffalo, NY 14260 \\
	\texttt{tianlema@buffalo.edu} \\
	\And
	Aidong Zhang\\
	Department of Computer Science\\
	University at Buffalo\\
	Buffalo, NY 14260 \\
	\texttt{azhang@buffalo.edu} \\}

\begin{document}

\maketitle

\begin{abstract}
While deep learning has achieved great success in many fields, one common criticism about deep learning is its lack of interpretability. In most cases, the hidden units in a deep neural network do not have a clear semantic meaning or correspond to any physical entities. However, model interpretability and explainability are crucial in many biomedical applications.
To address this challenge, we developed the Factor Graph Neural Network model that is interpretable and predictable by combining probabilistic graphical models with deep learning. We directly encode biological knowledge such as Gene Ontology as a factor graph into the model architecture, making the model transparent and interpretable. Furthermore, we devised an attention mechanism that can capture multi-scale hierarchical interactions among biological entities such as genes and Gene Ontology terms. With parameter sharing mechanism, the unrolled Factor Graph Neural Network model can be trained with stochastic depth and generalize well. 
We applied our model to two cancer genomic datasets to predict target clinical variables and achieved better results than other traditional machine learning and deep learning models. Our model can also be used for gene set enrichment analysis and selecting Gene Ontology terms that are important to target clinical variables.
\end{abstract}

\section{Introduction}
In computational biology, we often need to build predictive models using data-driven approaches. The predictors are usually a set of observable variables which we usually call features, and the targets are variables of interests we want to predict from the observable data. For instance, we may want to use gene expression data to predict disease status.

With a sufficient amount of data, machine learning especially deep learning models can achieve very high prediction accuracies. In fact, deep learning has already brought about breakthroughs in computer vision \citep{he2016deep}, speech recognition \citep{hinton2012deep}, natural language processing \citep{vaswani2017attention} and many other fields \citep{lecun2015deep}. However, conventional deep learning models require massive training data with clearly defined structure (such as images, audio, and natural languages), and are not directly suitable for many tasks in the biomedical domain.
Currently, deep learning is also insufficient to deal with the ``big p, small n'' problem (the number of features is large while the number of samples is relatively small) in many biomedical problems.

One common criticism about deep learning is its lack of interpretability \citep{gunning2017explainable}. In a deep neural network model, the hidden units are unknown and not interpretable. Intriguingly, neural network models with a different number of layers and hidden units can often generate very similar results.
Explainability and interpretability are crucial in many biomedical problems, for example, medical diagnoses. Developing predictable and explainable deep learning models \citep{ghosal2018explainable} is in urgent need.  

In this paper, we present the Factor Graph Neural Network (FGNN) model, which directly encodes biological knowledge such as Gene Ontology into the model architecture. Unlike the hidden nodes in conventional deep learning models which do not have a physical meaning, each node (i.e., ``neuron'') in the Factor Graph Neural Network model corresponds to some biological entity (such as genes or Gene Ontology terms), making the model transparent and interpretable. 
We not only address the model interpretability challenge but also make the model generalize well by directly incorporating biological knowledge as inductive biases \cite{battaglia2018relational} into the model architecture.
We also devised a parameter sharing mechanism to significantly reduce the number of model parameters and yet maintain the high representation power of deep learning models. Furthermore, we applied the attention mechanism to capture hierarchical multi-scale interactions between Gene Ontology terms and genes. Our model can be used for gene set enrichment analysis as well. Extensive experiments on two cancer genomic datasets demonstrated the effectiveness of the proposed model.

In the following, we briefly review some related work, then describe the proposed model followed by the experimental results. Finally, we conclude the paper with some discussion.
 
%Gene-gene interaction 
%
%Gene ontology:
%
%1) Each GO term consists of some genes
%2) Generalized gene set model
%3) Gene Ontology has hierarchical structure
%
%
%
%Every node correspond to some biological entity such as a gene, a pathway, or a GO term. 

%
%Incorporate prior knowledge as inductive bias

\section{Related work}
Factor graphs have been studied in probabilistic graphical models \citep{koller2009probabilistic}. Message-passing algorithms such as the sum-product algorithm are widely used for inference on factor graphs. Factor graphs have been used to infer patient-specific pathway activity with probabilistic inferences \citep{vaske2010inference}. These traditional algorithms are often not end-to-end differentiable and require a large amount of data to learn the model parameters. Recently, with the success of deep learning in many fields such as vision, language and game play \citep{lecun2015deep}, Bayesian deep learning combining Bayesian approaches and deep learning has drawn a lot of attention. For example, Variational AutoEncoder (VAE) \citep{kingma2013auto} used the reparameterization trick to enable gradient descent through random nodes in a neural network. However, the hidden nodes in the VAE model still lack a clear semantic meaning. 

As graphs are commonly used in many applications, novel deep learning models on graphs have been developed, including Graph Neural Network \citep{scarselli2009graph}, Graph Convolutional Neural Network \citep{kipf2016semi}, and so on. Graph Attention Model (GAM) \citep{velickovic2017graph} also applied the attention mechanism to learn graph embeddings. These approaches mainly focus on predictive tasks on a single graph, such as predicting node categories in a graph. By contrast, our proposed Factor Graph Neural Network model uses graphs from domain knowledge as model architecture and predicts clinical target variables.

Deep learning approaches had also been applied to sequencing data \citep{alipanahi2015predicting}, imaging data \citep{wang2016deep}, and medical records \citep{pham2016deepcare}. 
However, one drawback of most deep learning models is their lack of interpretability \citep{gunning2017explainable}. Recently, \citet{ghosal2018explainable} proposed an explainable deep machine vision framework for plant stress phenotyping by selecting the top-K high-resolution feature maps. However, this framework only works for vision-related convolutional neural network (CNN) models. For other fields, the feature maps in the convolutional layers are hardly interpretable. 
Our proposed Factor Graph Neural Network model is highly expressive and interpretable by combining the strength of interpretability of factor graphs in probabilistic graphical models and the supreme representation power of deep neural networks. 

%Linear model are often more interpretable, and can be combined with deep learning.
%Wide \& deep learning \citep{cheng2016wide} models are jointly trained wide linear models and deep neural networks and are used for recommender systems. However, they often cannot generalize well when the dataset is small.
%
%Denoising AutoEncoder (DAE) \citep{vincent2008extracting} makes the autoencoder model robust to partially corrupted input by training the model with randomly corrupted data. DAE can be used for data imputation \citep{gondara2018mida}.

Model generalizability is crucial in many real-world applications. Many regularization techniques have been introduced to deep learning models to make them generalize better. Dropout \citep{srivastava2014dropout} prevents overfitting by randomly setting part of the layer output to zero during training.
%AutoEncoder with multi-head attention \citep{vaswani2017attention} had been applied to sequence-to-sequence tasks successfully.
Relational inductive biases within deep learning model can facilitate interpretable relational reasoning \citep{battaglia2018relational}, and can also help the model generalize well.
\citet{huang2016deep} proposed to train a deep neural network with stochastic depth by randomly dropping out layers during training. This can be seen as an ``implicit'' ensemble of ResNet \citep{he2016deep} models and can generalize better than a single ResNet model with fixed depth. With parameter sharing mechanism, our proposed Factor Graph Neural Network model can also be trained with stochastic depth, which can further boost the model generalizability.

In computer vision, it is amazing that convolutional neural networks (CNN) with more than one thousand layers can still surpass human-level performance for image classification tasks \citep{huang2016deep}. One key to the success of CNN and other deep learning models is the use of proper relational inductive biases \citep{battaglia2018relational}. For CNN, the inductive biases include the locality (parameter sharing) and the translational invariance. 

To build predictable and generalizable deep learning models in the biomedical domain, we can incorporate some prior knowledge as inductive bias into the model, too. In this paper, our Factor Graph Neural Network model encodes the factor graph from domain knowledge as an inductive bias into the model architecture. It generalizes the Graph Convolutional Network (GCN) \citep{kipf2016semi} and can capture hierarchical multi-scale interactions with attention mechanisms.

%\citet{Steele2018} used data-driven approaches to select prognostic factors for predicting patient mortality in coronary artery disease, which outperformed traditional models using only a few expert-selected prognostic factors. 

%\section{Outline}
%Prototypical Network
%NN
%SVM, etc.
%
%Randomly assign edges
%
%Only use gene-gene interaction
%Only use gene-GO
%use gene-GO and GO hierarchy
%use gene-gene, gene-GO, and GO hierarchy
%
%
%Now prepare data

\section{Factor Graph Neural Network Model}
 
In many applications, there are two types of variables: observable variables and latent variables. Latent variables can be seen as ``factors'' that are related to observable variables. For example, gene expressions are measured in many genetic disease studies. These gene expressions are observable variables. A pathway or Gene Ontology (GO) term involves multiple genes and gene products, but their activities are not directly observable. These pathways and GO terms constitute various factors in a gene network. Since these hidden factors are not directly observable, many models directly use observable variables for predicting target variables such as clinical outcomes.

\begin{figure}
	\centering
	\includegraphics[width=0.8\linewidth]{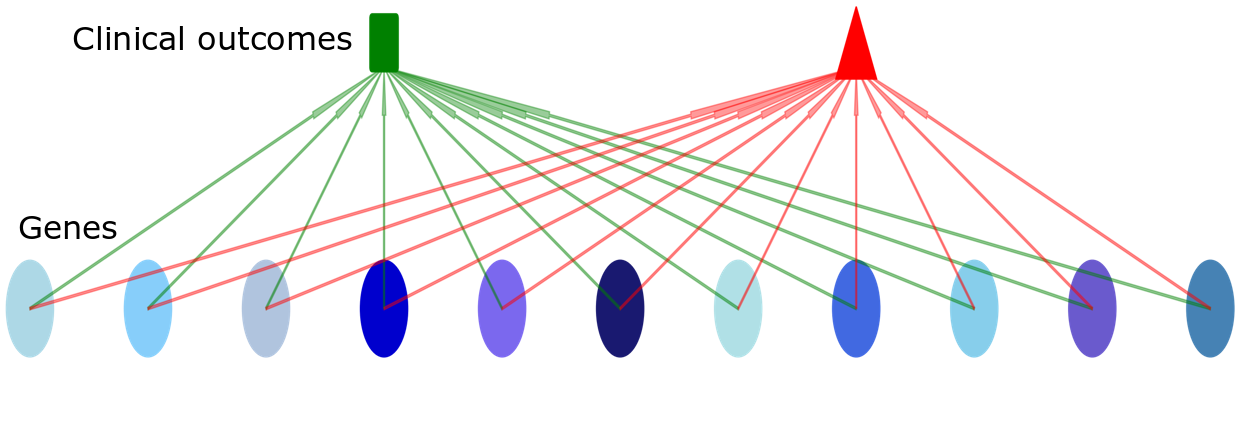}
	\caption{A simple ``shallow'' predictive model}
	\label{fig:one-layer}
\end{figure}

Fig.~\ref{fig:one-layer} shows a simple linear predictive model using the observable gene expressions $\mathbf{x}$ to predict the clinical outcomes $y$.
%In Fig.~\ref{fig:one-layer}, the input layer is the gene expressions $\mathbf{x}=(x_1, x_2, \cdots, x_n)$ and the output is the clinical outcomes $y$.

\begin{equation}\label{eq:simple-linear-model}
	y = \beta_{0} + \sum_{i=1}^{p} \beta_{i} * x_{i}
\end{equation}

In this simple linear model Eq.~\ref{eq:simple-linear-model}, $x_{1}, x_{2}, \cdots, x_{n}$ are the predictors, and $y$ is the target. We can apply more complex transformations to the predictors and to model the relationship between $\mathbf{x}$ and $y$. In order to learn the model parameters, we need to construct a dataset consisting of both predictors and targets and train the model with supervised learning. Note this model is ``shallow'', which can be seen as a one-layer neural network (we often do not count the input layer).

While shallow models such as generalized linear model and SVM have a limited representation power, they are still widely used in various applications, especially when the model interpretability and generalizability are crucial and the dataset is small.
On the other hand, deep neural network models have more representation power to approximate almost any complex nonlinear transformations. However, most deep learning models still lack interpretability and explainability.
Deep learning models usually have millions or even billions of parameters. We cannot assign semantic meanings to the hidden units and the parameters in a conventional neural network model. They can easily overfit almost any dataset with proper training, however, the model may not be able to generalize well on a new dataset. 

In order to make our Factor Graph Neural Network model predictable and generalizable, we incorporate prior knowledge such as Gene Ontology annotations as the inductive bias into the model architecture. Genes and Gene Ontology (GO) terms form a bipartite graph. There are two types of nodes (i.e., gene nodes and GO nodes). Each GO term is associated with a number of genes and gene products. GO terms are treated as factors in the Factor Graph Neural Network model. (Note we can also use pathways or other gene sets derived from biological knowledgebase as factors.) Based on Gene Ontology annotations, we can build a factor graph with GO terms as factors and genes as observable variables. This factor graph encodes domain knowledge and can be used as an inductive bias for constructing the Factor Graph Neural Network model.
%In the following, for ease of description, we choose gene and Gene Ontology (GO) terms as an example to describe the model. 

Suppose there are $k$ factors (i.e.,, GO terms) $F=\{f_1, f_2, \cdots, f_k\}$ and $n$ observable variables (i.e., genes) $X=\{x_1, x_2, \cdots, x_n\}$. Each factor $f_j$ has its domain on a subset of the observable variables $D_j \subset X$.

\begin{equation} \label{eq:variable-to-factor}
	f_j = \phi_j (D_j), \qquad j = 1, 2, \cdots, k
\end{equation}

$\phi_j$ in Eq.~\ref{eq:variable-to-factor} is an unknown complex function that maps the set of observable variables $D_j$ to factor $f_j$. We can use a neural network to approximate $\phi_j$.

\begin{figure}
	\centering
	\includegraphics[width=0.8\linewidth]{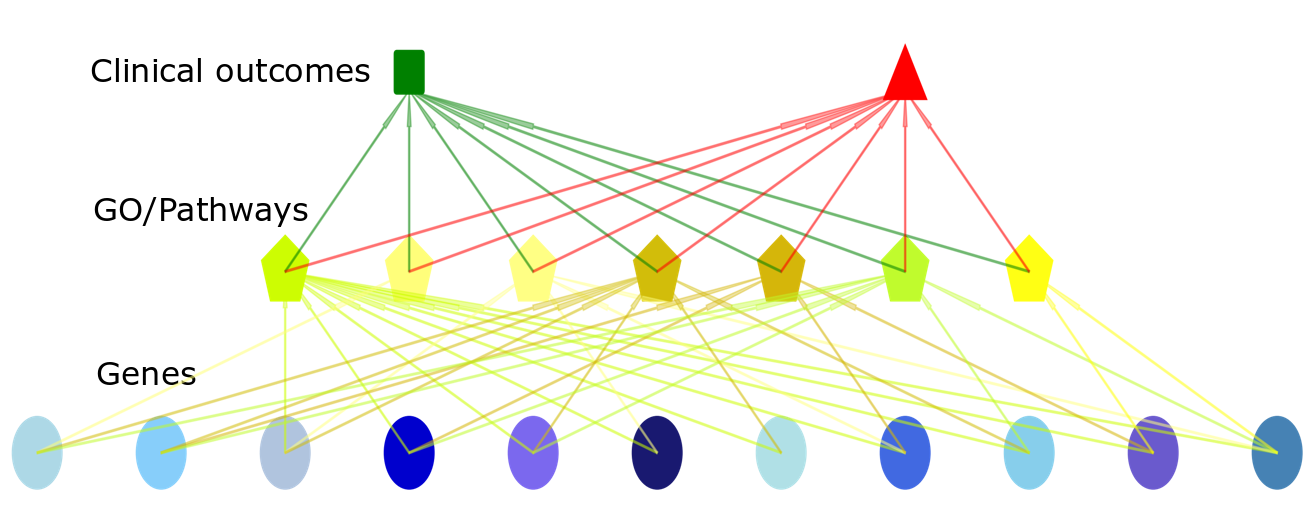}
	\caption{A simple factor graph neural network model}
	\label{fig:two-layer}
\end{figure}

Fig.~\ref{fig:two-layer} shows a simple two-layer Factor Graph Neural Network model. The nodes in the input layer are genes, and the nodes in the hidden layer are GO terms. The output layer is the target clinical outcomes. There is an edge between a gene and a GO term if and only if the gene is included in the GO term. Therefore the network is not fully connected between the input layer and the hidden layer. 
Instead, the connections are determined by the relationships between GO terms (factors) and genes (observable variables).

Each GO term is modeled by a small neural network model (Eq.~\ref{eq:variable-to-factor}). In total, we will have $k$ neural network models to approximate $k$ GO terms (factors). Note we have combined all the $k$ neural network models into one model with sparse connections in Fig.~\ref{fig:two-layer}. 
We can then use these $k$ factors to make predictions on the target clinical variable. 
The overall model can be seen as a network of networks. Each GO term (biological pathway) form a subnetwork, and all different GO terms (pathways) interact with each other forming a complex network of subnetworks.

\subsection{Unrolled Factor Graph Neural Network model}

The two-layer Factor Graph Neural Network model as shown in Fig.~\ref{fig:two-layer} is a two-layer neural network with its hidden layer corresponding to Gene Ontology terms and the edges determined by the Gene Ontology annotations. It is usually not sufficient for modeling complex nonlinear transformations with  a ``shallow'' architecture.
In order to make this model more expressive to model complex nonlinear transformations, we can unroll this Factor Graph Neural Network model to make it have more layers as deep neural networks can have more expressive power.

\begin{figure}
	\centering
	\includegraphics[width=0.8\linewidth]{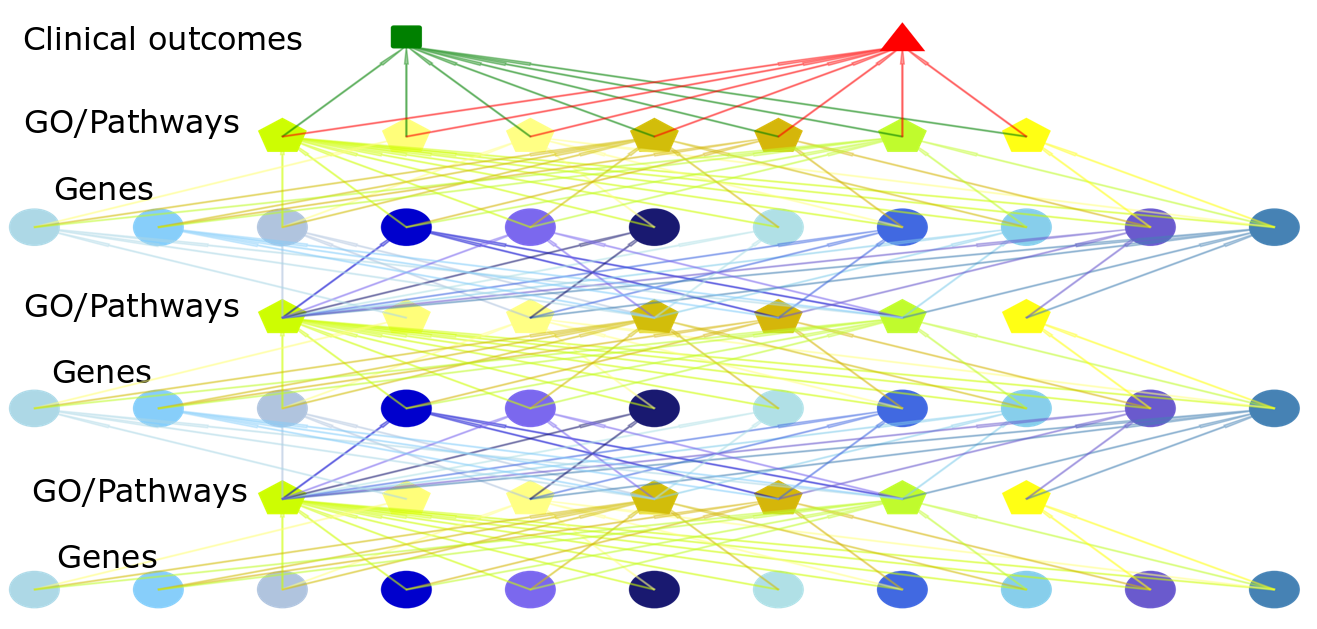}
	\caption{Unrolled factor graph neural network model}
	\label{fig:six-layer}
\end{figure}

Fig.~\ref{fig:six-layer} shows an unrolled factor graph neural network model with six layers. The input layer is the observable gene expressions $\mathbf{x}$. The first hidden layer corresponds to the GO terms. The second hidden layer corresponds to some latent state of genes. The rest of the hidden layers all correspond to latent states of genes or GO terms. The output layer is the target clinical variable $y$.

By unrolling the factor graph neural network model, we introduced multiple composable nonlinear transformations, making the model able to approximate any complex nonlinear functions. This unrolling operation is inspired by the efficient learning of deep Boltzmann machines \citep{salakhutdinov2010efficient}. 

\subsubsection{Parameter sharing among layers}
Though Fig.~\ref{fig:six-layer}  only shows a six-layer factor graph neural network model, there can be infinitely many layers by unrolling the model. If each layer has a different set of parameters as in widely used convolutional neural networks, then the number of parameters will be too large for most datasets in the biomedical domain which usually have the ``big p, small n'' problem. In order to reduce the risk of overfitting, we adopt a parameter sharing mechanism similar to recurrent neural networks and the Transformer model \citep{vaswani2017attention}. Different from convolutional neural networks where parameter sharing is within a single layer, our factor graph neural network model shares parameters across layers. 

There are two set of parameters in this unrolled factor graph neural network model: one used to map genes to GO terms, and the other map GO terms to genes. 
We have introduced the transformations $\phi_j$ from a subset of gene variables $D_j$ to GO factor $f_j$ (Eq.~\ref{eq:variable-to-factor}). We can also infer or reconstruct the hidden states of gene variables $\mathbf{X}$ from the GO factors $F$.

\begin{equation}
	x_i = \varphi_i(\mathcal{D}_i), \qquad i=1,2,\cdots, n
\end{equation}

$x_i$ is the $i$th variable (e.g., gene), while $\mathcal{D}_i \subset F$ is a subset of factors (GO terms) that include $x_i$.

In an unrolled factor graph neural network model with multiple layers, we can share these two set of parameters (corresponding to the two set of transformations from variables to factors and vice versa) across layers. As the same set of parameters are used between variables and factors in all the layers, we can significantly reduce the number of model parameters.
Since there can be multiple layers corresponding to genes and GO terms, we can add skip connections as in ResNet \citep{he2016deep} to connect the previous gene/GO layers to the current gene/GO layers. This can help gradient flow and speed up training.

\paragraph{Train the model with stochastic depth}
With parameter sharing in the unrolled factor graph neural network model, we can train the model with stochastic depth \citep{huang2016deep}. This can serve as an implicit regularizer similar to a Dropout layer.

\subsection{Attention mechanism for hierarchical multi-scale interactions}

Up to now we only considered immediate interactions among nodes in a factor graph, i.e., the direct connection between observable variables and factors. With unrolled factor graph neural network, we can naturally incorporate multi-scale hierarchical interactions into the model using attention mechanisms. 

To enable multi-scale hierarchical interaction in an unrolled factor graph neural network model, we connect the $l$th layer with all previous $(l-1)$ layers, similar to DenseNet \citep{huang2017densely}. However, we do not use dense connections between layers. Instead, we only connect nodes that can reach each other in the factor graph within a number of steps.

More specifically, the connections between $l$th layer and $(l-1)$th layer are simply the edges in the factor graph. There will be an edge between a node in the $l$th layer and a node in the $(l-1)$th layer if and only if node $i$ can reach node $j$ in the factor graph in one step. In other words, $j$ is the direct neighbor of $i$.
Similarly, there will be an edge between node $i$ in the $l$th layer and node $j$ in the $(l-k)$th layer if and only if node $i$ can reach node $j$ in $k$ steps. For example, when $k=2$, node $i$ and $j$ are connected through one neighbor, and all the connections capture neighbors of neighbors in the factor graph.

This idea is very much like convolutional neural networks (CNN) on graphs, where the neurons in the high-level layers will have a larger reception field. 
The difference is that in a CNN model, the number of neurons in each layer will be decreased by a factor of the stride, and the new feature plane no longer has a clear interpretable physical meaning. By contrast, every neuron in the factor graph neural network model corresponds to some physical entity (genes or GO terms), making the model transparent and interpretable. In the following, we present the algorithm to calculate the attention matrices that are used to connect all the layers in a factor graph neural network model to capture multi-scale hierarchical interactions.

\subsubsection{Attention matrices for capturing multi-scale hierarchical interactions}
Instead of simply connecting nodes across layers based on hierarchical interactions, we apply the attention mechanism to assign weights to connections between different layers.
The unrolled factor graph neural network model is based on the factor graph which encodes biological knowledge such as Gene Ontology annotations. (Note a factor graph is a data structure encoding biomedical domain knowledge, while the factor graph neural network model is a deep neural network model which uses the domain knowledge factor graph as the model backbone.)
A factor graph can be seen as an undirected bipartite graph. There are two node sets: source variables and target factors. Based on the network topology, we can calculate the state transition matrices from source to target and vice versa. One simplest state transition matrix is the normalized adjacency matrix. Algorithm~\ref{alg:attention_mats} shows how to calculate the attention matrices from the state transition matrices. Since the $l$th layer is connected to all previous $l-1$ layers in a factor graph neural network model, there are $l-1$ attention matrices for the $l$th layer used to attend each of the previous $l-1$ layers. 
These attention matrices provide with the weights for the connections across different layers, which capture multi-scale hierarchical interactions among nodes.
Note we can apply Algorithm~\ref{alg:attention_mats} to non-bipartite graphs as well by setting the source nodes and target nodes the same.

\begin{algorithm}
	\DontPrintSemicolon
	\SetKwInOut{Input}{Input}\SetKwInOut{Output}{Output}
	\Input{The state transition matrix from source variables to target factors of a factor graph: $M_s$\\
		The state transition matrix from target factors to source variables: $M_t$ (for undirected graph: $M_t=M_s^T$)\\ 
		Number of levels: $l$}
	\Output{A lists of attention matrices from the source variables to the nodes in all previous layers: $A_s$\\
		A lists of attention matrices from the factors to the nodes in all previous layers: $A_t$\\}
	\BlankLine
	\emph{$A_s=[M_s], A_t=[M_t]$}\\
	\For{$i\leftarrow 2$ \KwTo $l$}
	{
		\emph{$M_{s}^{'}=M_s, M_{t}^{'}=M_t$}\\
		\For{$j\leftarrow i-1$ \KwTo $1$}{
			\If(\tcp*[f]{Layer $i$ and $j$ correspond to the same node set}){$(i-j)\%2=0$}
			{$M_{s}^{'}=M_{s}^{'} \cdot M_s$\\ $M_{t}^{'}=M_{t}^{'} \cdot M_t$}
			\If(\tcp*[f]{$i$ and $j$ correspond to different node sets}){$(i-j)\%2 \ne 0$}
			{$M_{s}^{'}=M_{s}^{'} \cdot M_t$\\ $M_{t}^{'}=M_{t}^{'} \cdot M_s$}
		}
		$A_s=A_s + [M_{s}^{'^T}], A_t=A_t + [M_{t}^{'^T}]$
	}
	\caption{Generate attention matrices for a factor graph neural network model}\label{alg:attention_mats}
\end{algorithm} 

\subsubsection{Layer normalization}
Since we connect each layer with all the previous layers, the outputs from higher layers may become larger and larger. To ensure the outputs of all layers are on the same scale, we apply layer normalization before producing the final output of each layer.

\begin{equation}
\begin{split}
h_i &= \frac{h_i-\mu}{\sigma}\\
\mu &= \frac{1}{n} \sum_{i=1}^{n} h_i \\
\sigma^2 &= \frac{1}{n-1} \sum_{i=1}^{n} (h_i - \mu)^2
\end{split}
\end{equation}

Here, $\mu$ and $\sigma$ are layer mean and standard deviation. By applying layer normalization, we ensure that the output of each layer is on the same scale. This will stabilize training and make the learned representations for all the nodes (source variables or target factors) lie on the same manifold.

\subsubsection{Forward pass of the factor graph neural network model}
Since each factor has a different domain (reception field), we need to compute each factor separately. Sequential computation is time-consuming if the number of factors is large. With parameter sharing, we can parallelize the computation.

We use matrix $M_s$ to store all the weights used to transform from source variables (i.e., genes) to target factors (i.e., GO terms), and $M_t$ to store all the weights used to reconstruct source variables (or their hidden states) from the factors. 

Let $A_s$ and $A_t$ be the list of attention matrices generated by Algorithm~\ref{alg:attention_mats} from gene layer and GO layer to all the previous layers. 
The forward pass algorithm of an unrolled factor graph neural network model is shown in Algorithm~\ref{alg:bipartite_graph1d}.

\begin{algorithm}
	\DontPrintSemicolon
	\SetKwInOut{Input}{Input}\SetKwInOut{Output}{Output}
	\Input{Feature matrix $X$ of observable variables (source nodes) \\
		The weight matrix from source to target: $M_s$\\
		The weight matrix from target to source: $M_t$ (for undirected graph: $M_t=M_s^T$)\\
		A lists of attention matrix from the source nodes to the nodes in previous layers: $A_s$\\
		A lists of attention matrix from the target nodes to the nodes in previous layers: $A_t$\\ 
		Maximal number of layers: $M$\\
	Minimal number of layers: $N$}
	\Output{The outputs of each layer $H$}
	\BlankLine
	\emph{Randomly choose a number $l (N \le l \le M)$}\\
	\emph{$H = [X]$}\\
	\For{$i\leftarrow 2$ \KwTo $l$}
	{
		\emph{$Y=[\ ]$}\\
		\For{$j\leftarrow i-1$ \KwTo $1$}{
			\If(\tcp*[f]{Layer $i$ and $j$ are source nodes}){$i\%2=0\quad \& \quad j\%2=0$}
			{$y=H[j]\cdot A_s[i-j]$}
			\If(\tcp*[f]{Layer $i$ is target and $j$ is source}){$i\%2!=0\quad \& \quad j\%2=0$}
			{$y=H[j]\cdot M_s \cdot A_t[i-j]$}
			\If(\tcp*[f]{Layer $i$ and $j$ are target nodes}){$i\%2=0\quad \& \quad j\%2!=0$}
			{$y=H[j]\cdot A_t[i-j]$}
			\If(\tcp*[f]{Layer $i$ is source and $j$ is target}){$i\%2=0\quad \& \quad j\%2!=0$}
			{$y=H[j]\cdot M_t \cdot A_s[i-j]$}
			$Y = Y + [y]$
		}
		$H = H + [mean(Y)]$
	}
	\caption{Forward pass of the factor graph neural network model}\label{alg:bipartite_graph1d}
\end{algorithm}

In Algorithm~\ref{alg:bipartite_graph1d}, we randomly choose $l (N \le l \le M)$ as the number of layers for training the model with stochastic depth. We use attention matrices $A_s$ and $A_t$ to capture the weighted multi-scale interactions among nodes in the factor graph.

For supervised learning, we can add a classification head or regression head using the output of the last layer(s). We can use stochastic gradient descent to update model parameters $M_s$ and $M_t$. The whole framework is end-to-end differentiable.

\paragraph{Apply to interaction networks}
Our factor graph neural network model can be applied to non-bipartite graphs as well. For non-bipartite graphs, we can treat the source and target nodes as the same set of nodes and set model parameters $M_s=M_t$ in Algorithm~\ref{alg:bipartite_graph1d}. The whole framework can work seamlessly for non-bipartite graphs as well.

\paragraph{Encode network hierarchy}
If we have additional topological information about the factors such as Gene Ontology hierarchical structure, we can also encode the network hierarchy in the Factor Graph Neural Network model.
A hierarchical network such as Gene Ontology can be encoded as a list of $<child, parent, is-a>$ relations. For example GO:0000038 is a GO:0006631. (Here GO:0000038 is the child node and GO:0006631 is the parent node.)
This hierarchical network can be seen as a directed acyclic graph. Each node can be seen as a factor with its domain being all its children. Thus the factor graph neural network can be applied to encode network hierarchy as well.

%\section{GeneNet}
%
%first layer gene expression
%
%gene interaction
%
%gene pathway
%
%pathway hierarchy (add a dag layer)
%
%multi-scale network
%
%prediction

\paragraph{Factor Graph Neural Network for Gene Set Enrichment Analysis}
The factor graph neural network model can also be used for gene set enrichment analysis (GSEA).
Currently, gene set enrichment analysis mainly employs statistical approaches to identify the enriched gene sets. 
The factor graph neural network can be used to identify gene sets that are relevant to target clinical variables of interest based on stochastic gradient descent.

In fact, each GO term is associated with a set of genes and can thus be seen as a gene set. Our approach can be applied to identify enriched gene sets by training the model end-to-end. The weights of the last layer from the gene sets to the target clinical variable can be used to select gene sets that are most relevant to the target variable.

\section{Experiments}

\subsection{Dataset}
We downloaded the harmonized gene expression datasets for Lung Squamous Cell Carcinoma (project ID: TCGA-LUSC) and Kidney Renal Clear Cell Carcinoma (project ID: TCGA-KIRC) from Genomic Data Commons Data Portal (\url{https://portal.gdc.cancer.gov/}). First, we are trying to use gene expression profiles to predict tumor stage. We selected 246 Lung Squamous Cell Carcinoma primary solid tumor samples from two different tumor stages (``stage ib'': 152 samples, ``stage iib'': 94 samples), and 476 Kidney Renal Clear Cell Carcinoma primary solid tumor samples from three different stages (``stage i'': 271 samples, ``stage iv'': 82 samples, ``stage iii'': 123 samples). Other stages have too few samples and are thus being discarded. The tumor stage information was also retrieved from Genomic Data Commons Data Portal.

For our factor graph neural network model and Graph Convolutional Neural Network model, we also need Gene Ontology (GO) information. We downloaded and processed the current release of the human Gene Ontology Annotation file from \url{ftp://ftp.ebi.ac.uk/pub/databases/GO/goa/HUMAN/}.

\subsection{Data preprocessing}
For gene expression data (RNA-seq HTSeq counts), we performed log2 transformation and selected top 5000 most variant genes for downstream analysis. 
We normalized gene expression data to have mean equal to 0 and standard deviation equal to 1 for all samples.

After filtering out GO terms that have less than five genes in the selected gene set, 2597 GO terms had been selected as factors for the TCGA-LUSC project, and 2630 GO terms for the TCGA-KIRC project.

For the experiments, we randomly split the dataset into three sets: 70\% for training, 10\% for validation, and 20\% for testing.
We trained different models on the training set, and evaluated them on the validation set. We chose the model with the best validation accuracy to make predictions on the test set, and reported the precision, recall and F1 values on the test set. We shuffled the data and repeated the process ten times, and reported the average metrics for comparing model performances.

\begin{table}
	\caption{Results for Lung Squamous Cell Carcinoma dataset}
	\label{tbl:f1_LUSC}
	\centering
	\begin{tabular}{cccc}
		\toprule
		Model Name & Precision & Recall & F1 \\
		\midrule
		Random Forest & 0.518 & 0.547 & 0.520 \\
		Decision Tree & 0.529 & 0.524 & 0.522 \\
		MLP & 0.477 & 0.592 & 0.511 \\
		GCN & 0.512 & 0.596 & 0.522 \\
		Prototypical Network & 0.425 & 0.598 & 0.484   \\
%		Random Factor Graph & 0.568 & 0.588 & 0.539 \\
		Factor Graph Neural Network & \textbf{0.557} & \textbf{0.624} & \textbf{0.566}\\
		\bottomrule
	\end{tabular}
\end{table}

\begin{table}
	\caption{Results for Kidney Renal Clear Cell Carcinoma dataset}
	\label{tbl:f1_KIRC}
	\centering
	\begin{tabular}{cccc}
		\toprule
		Model Name & Precision & Recall & F1 \\
		\midrule
		Random Forest & 0.525 & 0.594 & 0.532 \\
		Decision Tree & 0.495 & 0.518 & 0.500  \\
		MLP & 0.427 & 0.587 & 0.482 \\
		GCN & 0.519 & 0.603 & 0.533  \\
		Prototypical Network & 0.341 & 0.566 & 0.422 \\
%		Random Factor Graph & 0.525 & 0.602 & 0.532 \\
		Factor Graph Neural Network & \textbf{0.562} & \textbf{0.603} & \textbf{0.563}\\
		\bottomrule
	\end{tabular}
\end{table}

We compared our methods with traditional machine learning methods Random Forest, Decision Tree and Multilayer Perceptron (MLP), as well as two recent deep learning methods including Graph Convolutional Neural Network (GCN) \cite{kipf2016semi}, and Prototypical Network \cite{snell2017prototypical}.

Table~\ref{tbl:f1_LUSC} and Table~\ref{tbl:f1_KIRC} showed the experimental results on the lung cancer and kidney cancer datasets. On both datasets, the proposed factor graph neural network model achieved the best precision, recall and F1 scores.

In order to see the contribution of incorporating Gene Ontology into the model, we randomly assign genes to gene sets and build a random factor graph for the model. The model performed much worse after randomization on both dataset: the F1 score dropped from 0.566 to 0.539
on the lung cancer dataset and from 0.563 to 0.532 on the kidney cancer dataset. 

\subsection{Predicting tumor-normal sample type}
We also used our proposed method to classify sample types (primary solid tumor and solid tissue normal) on Kidney Renal Clear Cell Carcinoma dataset which has 538 tumor samples and 72 normal samples. By comparing the tumor samples with the normal samples, we can get insights about the molecular underpinning of tumors. 

As normal samples and tumor samples are very different and are almost linearly separable, we used 5\% of the data (29 samples) as the training set, 5\% (29 samples) as the validation set, and the rest 90\% (552 samples) as the test set. Both the Factor Graph Neural Network and Multilayer Perceptron (MLP) achieved 98\% accuracy on the test set. However, our Factor Graph Neural Network model is interpretable while MLP and other models are not.

\begin{figure}
	\centering
	\begin{subfigure}[b]{0.5\textwidth}
		\includegraphics[width=\textwidth]{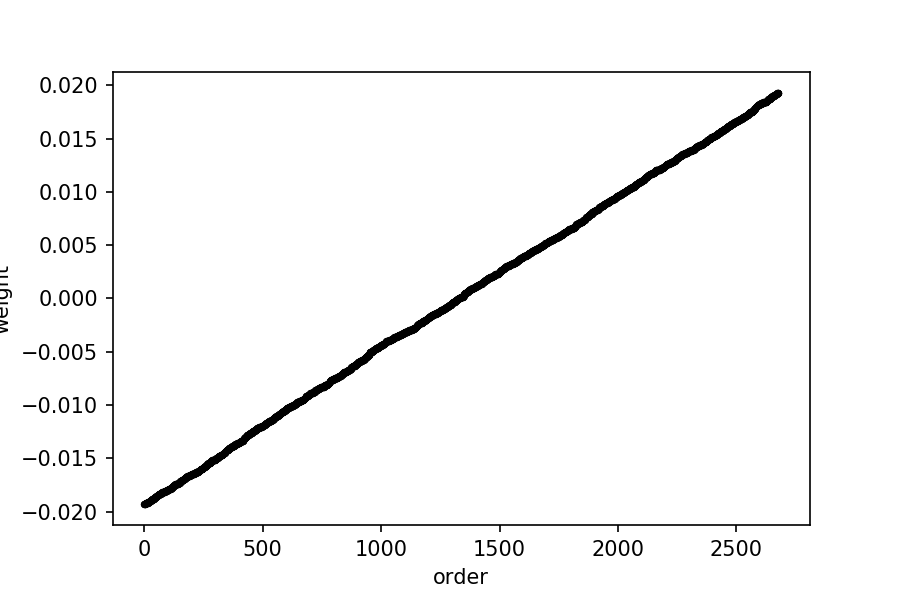}
		\caption{GO weight \textbf{before} training}
		\label{fig:random_weight}
	\end{subfigure}
	~ 
	\centering
	\begin{subfigure}[b]{0.5\textwidth}
		\includegraphics[width=\textwidth]{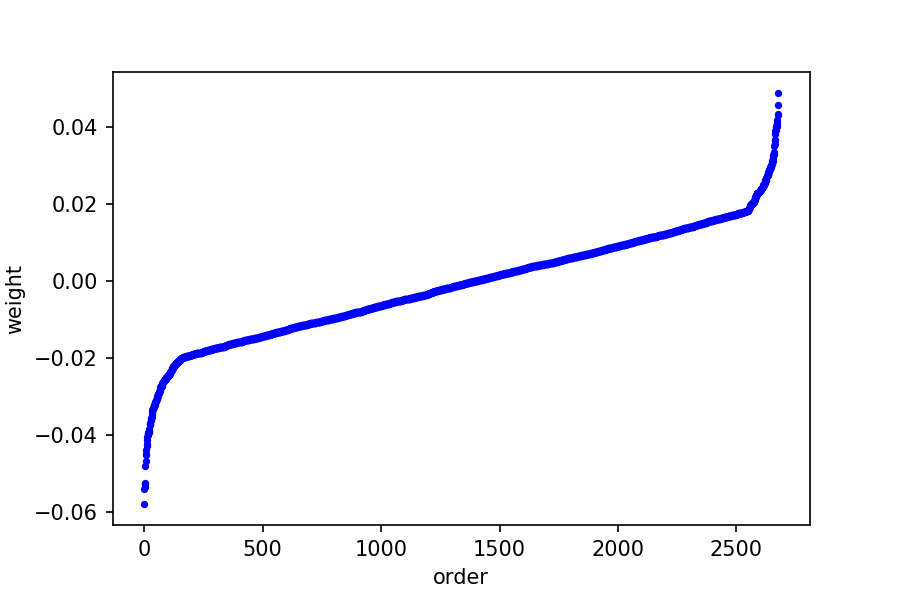}
		\caption{GO weight \textbf{after} training}
		\label{fig:learned_weight}
	\end{subfigure}
	\caption{GO weight before and after training}\label{fig:go_weight_KIRC}
\end{figure}

In the Factor Graph Neural Network model, the last layer is a linear classifier. The input of the classifier is the learned representations of Gene Ontology terms. The weight matrix of the linear classifier can be used to select Gene Ontology terms that are important to distinguish tumor and sample types.

We plotted the weights for 2678 GO terms in the last hidden layer of the model. Before training the model parameters are randomly initialized as shown in Figure.~\ref{fig:random_weight}. After training, most GO terms have a small weight in absolute value, while only a few have high weights as shown in Figure.~\ref{fig:learned_weight}. Table~\ref{tbl:top_GO_KIRC} shows the top ten GO terms with the highest absolute weights learned by the factor graph neural network model. The relevance of these GO terms to the kidney cancer can be verified by the domain experts.

\begin{table}
	\caption{GO terms with highest (absolute) weight for Kidney Renal Clear Cell Carcinoma}
	\label{tbl:top_GO_KIRC}
	\centering
	\begin{tabular}{ccc}
		\toprule
		GO ID & GO Name & Weight \\
		\midrule
		GO:0030574 & C-type mannose receptor 2 & 0.058 \\
		GO:0001556 & YTH domain-containing family protein 2 & 0.054 \\
		GO:0021542 & Nuclear receptor subfamily 2 group E member 1 & 0.054 \\
		GO:0007267 & Tumor necrosis factor receptor superfamily member 11A & 0.053 \\
		GO:0005813 & Coiled-coil domain-containing protein 61 & 0.052 \\
		GO:0052695 & UDP-glucuronosyltransferase 2A1 & 0.049 \\
		GO:0042359 & Legumain & 0.048 \\
		GO:0007254 & NF-kappa-B essential modulator & 0.047 \\
		GO:0047086 & Putative aldo-keto reductase family 1 member C8 & 0.046 \\
		GO:0048387 & Cytochrome P450 26B1 & 0.045 \\
		\bottomrule
	\end{tabular}
\end{table}

\section{Conclusion}
While data is abundant in many domains such as text, images, videos, etc., biomedical data including genomic data is usually not sufficient for large-scale machine learning.
In order to build a predictable and generalizable deep learning model, we usually need to incorporate some biological knowledge as an inductive bias into the model. 

In this paper, we presented the Factor Graph Neural Network model. The model architecture is based on biomedical knowledge (i.e., Gene Ontology annotations). Each node in the Factor Graph Neural Network model corresponds to some biological entity such as genes or Gene Ontology terms, making the model transparent and interpretable.

In order to make the model expressive enough to capture any complex nonlinear relationships, we can unroll the Factor Graph Neural Network model to have infinitely many layers. With parameter sharing, the model can be trained with stochastic depth which can help the model generalize better. We also devised an attention mechanism to capture the multi-scale interactions among biological entities such as genes and Gene Ontologies terms. 

The experimental results on two cancer genomic datasets show that our proposed model outperformed other methods including Graph Convolutional Network \citep{kipf2016semi}. Though we mainly focused our discussion on biomedical data analysis, as a general framework, the Factor Graph Neural Network can be applied to any other data with a graph as prior knowledge for interpretable deep learning. 
%The ultimate goal of omic data analysis is to disentangle complex factors and identify important factors that contribute to disease etiology. Our model is able to learn a distributed representations of molecular entities and patients and facilitates mining relationships among molecular features and clinical features. Essentially, learning a good representation of both molecular and clinical features is fundamentally important to unravel the intricate relationships among them. Our work provides a proof-of-concept framework for unifying data-driven and knowledge-driven approaches for mining multi-omic data with biological knowledges. We hope it can be applied to large-scale cancer genomic data and can contribute to elucidating the etiology and mechanisms of cancer and other complex diseases.

\bibliography{factor_graph.bib}
\bibliographystyle{icml2018}

\end{document}